\input{aipcheck}

\documentclass[
    ,final            % use final for the camera ready runs
  ]
  {aipproc}

\layoutstyle{6x9}

\begin{document}

\title{Hadron physics: a quark-model analysis}



\author{A. Valcarce}{
  address={Dpto. de F\'\i sica Fundamental, Universidad de Salamanca, Spain}
}

\author{J. Vijande}{
  address={Dpto. de F\'\i sica Te\'orica, Universidad de Valencia (UV) 
and IFIC (UV-CSIC), Spain}
}

\author{P. Gonz\'alez}{
  address={Dpto. de F\'\i sica Te\'orica, Universidad de Valencia, Spain}
}
\author{H. Garcilazo}{
  address={Escuela Superior de F\'\i sica y Matem\'aticas, Instituto Polit\'ecnico Nacional, Mexico}
}

\begin{abstract}
We discuss recent results on heavy and light baryon spectroscopy.
\end{abstract}

\maketitle

\section{Introduction}
Hadron spectroscopy has undergone a great renaissance in recent years~\cite{Ros07}. The
new findings include: low-lying excitations of D and B mesons, long-awaited missing
states and new states near 4 GeV/c$^2$ in the charmonium spectrum, charmed
and bottom baryons, and evidence for doubly charmed baryons. The light hadron
sector remains also restless reporting new scalar mesons or showing a deep theoretical
interest in the high energy part of both the meson and baryon spectra.
In this talk we center our attention in the heavy baryon spectroscopy as well as
some anomalous states present in the light baryon spectra.

\section{Heavy baryons}

Heavy baryons containing a single heavy quark are particularly
interesting. The light degrees of freedom (quarks and gluons)
circle around the nearly static heavy quark. Such a system behaves as
the QCD analogue of the familiar hydrogen bounded by the 
electromagnetic interaction. 
When the heavy quark mass $m_Q \to \infty$, the angular
momentum of the light degrees of freedom is a good 
quantum number. Thus, heavy quark baryons belong to 
either SU(3) antisymmetric $\mathbf{\bar{3}_F}$ or symmetric 
$\mathbf{6_F}$ representations. The spin of the light
diquark is 0 for $\mathbf{\bar{3}_F}$, while it is 1
for $\mathbf{6_F}$. Thus, the spin of the ground
state baryons is $1/2$ for $\mathbf{\bar{3}_F}$, representing
the $\Lambda_b$ and $\Xi_b$ baryons, while it can be
both $1/2$ or $3/2$ for $\mathbf{6_F}$, allocating
$\Sigma_b$, $\Sigma^*_b$, $\Xi'_b$, $\Xi^*_b$, $\Omega_b$
and $\Omega^*_b$, where the star indicates spin $3/2$ states.
Therefore heavy hadrons form doublets.
For example, $\Sigma_b$ and $\Sigma_b^*$ will be degenerate
in the heavy quark limit. Their mass splitting is caused
by the chromomagnetic interaction at the order $1/m_Q$.
These effects can be, for example, taken into account systematically in the
framework of heavy quark effective field theory.
The mass difference between states belonging to the 
$\mathbf{\bar{3}_F}$ and $\mathbf{6_F}$ representations
do also contain the dynamics of the light diquark 
subsystem, hard to accommodate in any heavy quark mass
expansion. Therefore, exact solutions of the three-body
problem for heavy hadrons are theoretically
desirable because they will serve to test the reliability   
of approximate techniques: heavy quark mass expansions, 
variational calculations, or quark-diquark approximations. 
\begin{table}
\begin{tabular}{|cc||ccccc|ccccc|}
\hline
\multicolumn{2}{|c||}{} &
\multicolumn{5}{|c|}{Charm} &
\multicolumn{5}{|c|}{Bottom} \\
State & $J^P$ & CQC & Exp.  & 
\protect\cite{Sil96} &
\protect\cite{Rob07} &
\protect\cite{Ebe08} & 
CQC & 
Exp.  & 
\protect\cite{Sil96} &
\protect\cite{Rob07} &
\protect\cite{Ebe08} \\
\hline
$\Lambda_i$  &  $1/2^+$ & 2285 & 2286 & 2285 & 2268 & 2297  
                        & 5624 & 5624 & 5638 & 5612 & 5622 \\
             &  $1/2^+$ & 2785 & 2765 & 2865 & 2791 & 2772 
                        & 6106 &      & 6188 & 6107 & 6086 \\
             &  $1/2^-$ & 2627 & 2595 & 2635 & 2625 & 2598  
                        & 5947 &      & 5978 & 5939 & 5930 \\
             &  $1/2^-$ & 2880 & 2880 & 2885 & 2816 & 3017  
                        & 6245 &      & 6268 & 6180 & 6328 \\
             &  $3/2^+$ & 3061 &      & 2930 & 2887 & 2874  
                        & 6388 &      & 6248 & 6181 & 6189 \\
             &  $3/2^+$ & 3308 &      & 3160 & 3073 & 3262  
                        & 6637 &      & 6488 & 6401 & 6540 \\
             &  $5/2^+$ & 2888 & 2880 & 2930 & 2887 & 2883  
                        &      &      &      &      &       \\ \hline
$\Sigma_i$   &  $1/2^+$ & 2435 & 2454 & 2455 & 2455 & 2439  
                        & 5807 & 5808 & 5845 & 5833 & 5805 \\
             &  $1/2^+$ & 2904 &      & 3025 & 2958 & 2864  
                        & 6247 &      & 6370 & 6294 & 6202 \\    
             &  $1/2^-$ & 2772 & 2765 & 2805 & 2748 & 2795  
                        & 6103 &      & 6155 & 6099 & 6108 \\
             &  $1/2^-$ & 2893 &      & 2885 &      & 3176  
                        & 6241 &      & 6245 &      & 6401 \\
             &  $3/2^+$ & 2502 & 2518 & 2535 & 2519 & 2518  
                        & 5829 & 5829 & 5875 & 5858 & 5834 \\
             &  $3/2^+$ & 2944 & 2940 & 3065 & 2995 & 2912  
                        & 6260 &      & 6385 & 6308 & 6222 \\ 
             &  $3/2^-$ & 2772 & 2800 & 2805 & 2763 & 2761  
                        &      &      &      &      &      \\ \hline          
$\Xi_i$      &  $1/2^+$ & 2471 & 2471 & 2467 & 2492 & 2481  
                        & 5801 & 5793 & 5806 & 5844 & 5812 \\
             &  $1/2^+$ & 3137 & 3123 & 2992 &      & 2923  
                        & 6258 &      & 6306 &      & 6264 \\
             & $1/2'^+$ & 2574 & 2578 & 2567 & 2592 & 2578  
                        & 5939 &      & 5941 & 5958 & 5937 \\
             & $1/2'^+$ & 3212 &      & 3087 &      & 2984  
                        & 6360 &      & 6416 &      & 6327 \\
             &  $1/2^-$ & 2799 & 2792 & 2792 & 2763 & 2801  
                        & 6109 &      & 6116 & 6108 & 6119 \\
             &  $1/2^-$ & 2902 &      & 2897 & 2859 & 2928  
                        & 6223 &      & 6236 & 6192 & 6238 \\
             &  $1/2^-$ & 3004 & 2980 & 2993 &      & 3186  
                        &      &      &      &      &      \\
             &  $3/2^+$ & 2642 & 2646 & 2647 & 2650 & 2654  
                        & 5961 &      & 5971 & 5982 & 5963 \\
             &  $3/2^+$ & 3071 & 3076 & 3057 & 2984 & 3030  
                        & 6373 &      & 6356 & 6294 & 6341 \\
             &  $5/2^+$ & 3049 & 3055 & 3057 &      & 3042 
                        &      &      &      &      &      \\
             & $5/2'^+$ & 3132 & 3123 & 3167 &      & 3123 
                        &      &      &      &      &       \\ \hline                        
$\Omega_i$   &  $1/2^+$ & 2699 & 2698 & 2675 & 2718 & 2698  
                        & 6056 &      & 6034 & 6081 & 6065 \\
             &  $1/2^+$ & 3159 &      & 3195 & 3152 & 3065 
                        & 6479 &      & 6504 & 6472 & 6440 \\
             &  $1/2^-$ & 3035 &      & 3005 & 2977 & 3020  
                        & 6340 &      & 6319 & 6301 & 6352 \\
             &  $1/2^-$ & 3125 &      & 3075 &      & 3371  
                        & 6458 &      & 6414 &      & 6624 \\
             &  $3/2^+$ & 2767 & 2768 & 2750 & 2776 & 2768 
                        & 6079 &      & 6069 & 6102 & 6088 \\
             &  $3/2^+$ & 3202 &      & 3235 & 3190 & 3119 
                        & 6493 &      & 6519 & 6478 & 6518 \\ \hline
\end{tabular}
\caption{Masses, in MeV, of charmed and bottom baryons.}
\label{t12}
\end{table}

We have solved the Schr\"odinger equation by the 
Faddeev method in momentum space with the constituent
quark model (CQC) of Ref.~\cite{PRD08}.
The results are shown in Table \ref{t12} compared to 
experiment and other theoretical approaches. 
All known experimental data are nicely described. 
Such an agreement and the
exact method used to solve the three-body problem make our
predictions also valuable as a guideline to experimentalists.

As compared to other results
in the literature we see an overall agreement for the low-lying
states both with the quark-diquark approximation of 
Ref.~\cite{Ebe08} and the variational calculation in a
harmonic oscillator basis of Ref.~\cite{Rob07}. 
It is worth noticing that the relativistic quark-diquark approximation
and the harmonic oscillator variational method predict a lower
$3/2^+$ excited state for the $\Lambda_b$ baryon.
Such result can be easily understood by looking at Table~\ref{t7},
where it is made manifest the influence  of the pseudoscalar 
interaction between the light quarks on the 
$\Lambda_b(1/2^+)$ ground state, diminishing its mass by 
200 MeV. If this attraction were not present for the $\Lambda_b(1/2^+)$, the 
$\Lambda_b(3/2^+)$ it would be lower in mass as reported
in Refs.~\cite{Rob07,Ebe08} (a similar effect will be 
observed in the charmed baryon spectra).
Thus, the measurement and identification of the 
$\Lambda_b(3/2^+)$ is a relevant feature that will help to
clarify the nature of the interaction between the light
quarks in heavy baryon spectroscopy, determining the need
of pseudoscalar forces consequence of the spontaneous
chiral symmetry breaking in the light flavor sector.

\begin{table}
\begin{tabular}{|c||ccc|}
\hline
State & 
Full & 
$V_\pi=0$ & 
$\Delta E$  \\
\hline
$\Sigma_b(1/2^+)$     & 5807 & 5822  & $-$15   \\
$\Sigma_b(3/2^+)$     & 5829 & 5844  & $-$15   \\
$\Lambda_b(1/2^+)$    & 5624 & 5819  & $-$195   \\
$\Lambda_b(3/2^+)$    & 6388 & 6387  & $+$ 1   \\ \hline
\end{tabular}
\caption{Masses, in MeV, of different bottom baryons with two-light
quarks with (Full) and without ($V_\pi=0$) the contribution of the one-pion exchange
potential.}
\label{t7}
\end{table}
In the case of charmed baryons, there are some excited states that it is
not even known if they are excitations of the $\Lambda_c$ or $\Sigma_c$.
Besides, a number of new $\Xi_c$ and $\Xi'_c$ states have been also 
discovered recently~\cite{PRD08}. As can be seen all known experimental states
fit nicely into the description of our model not leaving too many
possibilities open for the assigned quantum numbers as we resume
in Table~\ref{t30}.

\begin{table}[b]
\begin{tabular}{|c|c|}
\hline
Experimental resonance (MeV) & 
Model states \\ \hline
\multicolumn{2}{|c|}{ $\Lambda_c$ or $\Sigma_c$} \\ \hline
2765                         & $\Sigma_c(1/2^-)$ or ${\Lambda_c(1/2^+)}^*$  \\
2880                         & ${\Lambda_c(1/2^-)}^*$ or $\Lambda_c(5/2^+)$ \\
2940                         & ${\Sigma_c(3/2^+)}^*$                      \\
2800                         & $\Sigma_c(3/2^-)$                          \\ \hline
\multicolumn{2}{|c|}{ $\Xi_c$ or $\Xi'_c$} \\ \hline
3055                         & $\Xi_c(5/2^+)$ \\ 
3123			     & ${\Xi_c(1/2^+)}^*$ or $\Xi'_c(5/2^+)$ \\
2980                         & ${\Xi_c(1/2^-)}^*$ \\
3076                         & ${\Xi_c(3/2^+)}^*$ \\ \hline
\end{tabular}
\caption{Possible model states and spin-parity assignments
for recently discovered charmed baryons. The 'star' indicates
radial excitations.}
\label{t30}
\end{table}

Finally, we can make parameter free predictions
for ground states as well as for spin, orbital and radial excitation
of doubly charmed and bottom baryons. 
Our results are shown in Table~\ref{t20}.
For doubly charmed baryons,
the ground state is found to be at 3579 MeV, far below the result of
Ref.~\cite{Rob07} and in perfect agreement with lattice nonrelativistic
QCD~\cite{Mat02}, but still a little bit higher than the 
non-confirmed SELEX result, 3519 MeV~\cite{Och05}. It is therefore
a challenge for experimentalists to confirm or to find the ground state
of doubly charmed and bottom baryons. 
\begin{table}
\begin{tabular}{|cc|ccccc|}
\hline
State  & 
$J^P$ & 
CQC & 
\protect\cite{Ron95} & 
\protect\cite{Sil96} & 
\protect\cite{Mat02} & 
\protect\cite{Rob07} \\  
\hline
$\Xi_{bb}$      & $1/2^+$     &10189&10340&10194&     & 10340 \\ \cline{2-7}
                &             &     &     &     &     &        \\ 
                & $3/2^+$     &  29 &  30 &  41 &  20 &  27  \\  
                & ${3/2^+}^*$ & 312 &     & 386 &     & 238  \\   
$\Delta E$      & ${1/2^+}^*$ & 293 &     & 355 &     & 236  \\  
                & $1/2^-$     & 217 &     & 262 &     & 153  \\  
                & ${1/2^-}^*$ & 423 &     & 462 &     & 370  \\  \hline
$\Omega_{bb}$   & $1/2^+$     &10293&10370&10267&     & 10454  \\\cline{2-7}
                &             &     &     &     &     &        \\ 
                & $3/2^+$     &  28 &  30 &  38 &  19 &  32  \\ 
                & ${3/2^+}^*$ & 329 &     & 383 &     & 267  \\   
$\Delta E$      & ${1/2^+}^*$ & 311 &     & 359 &     & 239  \\  
                & $1/2^-$     & 226 &     & 265 &     & 162  \\  
                & ${1/2^-}^*$ & 390 &     & 410 &     & 309  \\  \hline
$\Xi_{cc}$      & $1/2^+$     & 3579& 3660& 3607& 3588& 3676  \\  \cline{2-7}
                &             &     &     &     &     &        \\        
                & $3/2^+$     &  77 &  80 &  93 &  70 &  77  \\ 
                & ${3/2^+}^*$ & 446 &     & 486 &     & 366  \\   
$\Delta E$      & ${1/2^+}^*$ & 397 &     & 435 &     & 353  \\  
                & $1/2^-$     & 301 &     & 314 &     & 234  \\
                & ${1/2^-}^*$ & 439 &     & 472 &     & 398  \\  \hline
$\Omega_{cc}$   & $1/2^+$     & 3697& 3740& 3710& 3698& 3815  \\  \cline{2-7}
                &             &     &     &     &     &        \\            
                & $3/2^+$     &  72 &  40 &  83 &  63 &  61  \\
                & ${3/2^+}^*$ & 463 &     & 498 &     & 373  \\    
$\Delta E$      & ${1/2^+}^*$ & 415 &     & 445 &     & 365  \\  
                & $1/2^-$     & 312 &     & 317 &     & 231  \\ 
                & ${1/2^-}^*$ & 404 &     & 410 &     & 320  \\  \hline
\end{tabular}
\caption{Ground state and excitation energies, $\Delta E$, of doubly charmed and bottom
baryons. The 'star' indicates
radial excitations. Masses are in MeV.}
\label{t20}
\end{table}
  
The combined study of $Qqq$ and $QQq$ systems, where $Q$ stands for a heavy
$c$ or $b$ quark and $q$ for a light $u$, $d$, or $s$ quark, will also provide 
some hints to learn about the basic dynamics governing the interaction
between light quarks. The interaction between
pairs of quarks containing a heavy quark $Q$ is driven by the 
perturbative one-gluon exchange. For the $Qqq$
system the mass difference between members of the  
$\mathbf{6_F}$ SU(3) representation comes determined only
by the perturbative one-gluon exchange, whether between members
of the $\mathbf{6_F}$ and $\mathbf{\bar{3}_F}$ representations
it presents contributions from the one-gluon exchange and also
possible pseudoscalar exchanges. If the latter mass difference
is attributed only to the one-gluon exchange (this would be the case
of models based only on the perturbative one-gluon exchange), it will be strengthened
as compared to models considering pseudoscalar potentials at the 
level of quarks, where a weaker one-gluon exchange will play the role.
When moving to the $QQq$ systems only one-gluon exchange interactions
between the quarks will survive, with the strength determined in the
$Qqq$ sector, where we have experimental data. This will give rise
to larger masses for the ground states, due to the more attractive 
one-gluon exchange potential in the $Qqq$ sector, what requires
larger constituent quark masses to reproduce the experimental data.
This could be the reason for the larger masses of ground state doubly
heavy baryons obtained with gluon-based interacting 
potentials~\cite{Rob07,Cap86}. 

\section{Light baryons}

In the Particle Data Group (PDG) book~\cite{PDG06} the light-quark ($u$ and $%
d$) baryon spectrum is composed of forty resonances rated from one ($\ast )$
to four $(\ast \ast \ast \ast )$ stars. The PDG average--mass region below
1950 MeV contains mostly four--star (well established) resonances, fourteen
out of twenty three, the same being true for the $\Lambda$ strange sector,
eight out of eleven. This makes this mass region the most suitable for
testing any spectroscopic quark model. From the pioneering Isgur and Karl's
non-relativistic quark model in the late 70's~\cite{IK78} more refined
spectroscopic quark models for baryons, based on two-body interactions, have
been developed~\cite{PRC08}. We will
refer to them as two-body quark models and we shall denote them generically
as $3q^{2b}$. As an overall result the masses of the fourteen four-star
resonances, most times with the exception of $N_{P_{11}}(1440)$, are rather well predicted ($\leq 100$ MeV difference
with the PDG average value) by these models. Regarding the five three-star
(likely to certain existence) resonances, the situation is much less
favorable since the masses of two of them, $\Delta _{P_{33}}(1600)$ and $%
\Delta _{D_{35}}(1930),$ are generally overpredicted, up to 250 MeV above
the PDG average value. Let us note that a similar discrepancy is observed
for $\Delta _{S_{31}}(1900)(\ast \ast )$ and $\Delta _{D_{33}}(1940)(\ast )$
($\geq 100$ MeV difference with the PDG average value) which can be
related to $\Delta _{D_{35}}(1930)$ as we shall show, and for $\Delta
_{P_{31}}(1750)(\ast )$ (up to 200 MeV above the PDG average). In the
strange $\Lambda $ sector an outstanding overpredicted (by $80-150$ MeV)
state is the $\Lambda _{S_{01}}(1405)(\ast \ast \ast \ast )$. Henceforth we
shall call anomalies these significantly overpredicted mass resonances.

We carry out a general analysis of the anomalies: we
identify them and we propose a plausible physical mechanism to give
correctly account of their masses. 
Among the anomalies we find large-energy-step anomalies,
that correspond either to radial excitations as the $\Delta _{P_{33}}(1600)$
and the $N_{P_{11}}(1440)$ or quark Pauli blocking induced states as the $\Delta
_{D_{35}}(1930)$ and the $\Delta _{P_{31}}(1750)$~\cite{PRC08}.

Given the large radial excitation energy and the large mass predicted for
quark Pauli blocking induced states, one may wonder about the possibility
that $4q1\overline{q}$ components may be energetically competitive, despite
the extra quark and antiquark masses. Thus, they could greatly contribute,
altogether with $3q$ components, to the formation of the bound structures.
In order to examine this possibility at a phenomenological level we look for $%
4q1\overline{q}$ components in the form of inelastic meson-baryon channels
in relative $S$ wave (the lowest energy partial wave) with adequate quantum
numbers to couple to the anomalies and with thresholds close above their PDG
masses. We shall name these components meson-baryon threshold channels or $%
mB $ channels.

Certainly meson-baryon channel coupling effects may be at work for other
resonances not involving either large energy excitation steps or a large
mass induced by quark Pauli blocking. The most prominent examples are
the $\Lambda _{S_{01}}(1405)$ being mostly
interpreted, at the hadron level, as an $S$ wave $N\overline{K}$ quasi-bound
system, and the $\Delta _{F_{35}}(2000)$, a bizarre state since its 
average mass is obtained from three different data
analyses, two of them~\cite{Vra00} reporting a mass about $1720$ $(\pm
60)$ MeV and the other~\cite{Cut80} giving a quite different value of $%
2200\pm 125$ MeV. Then by considering two differentiated resonances the $%
\Delta _{F_{35}}(\sim 1720)$ would be a clear candidate for an anomaly.

To go beyond the qualitative analysis of the anomalies we
shall consider a system of one confined channel, the $3q^{2b}$, in
interaction with one free-channel, a meson-baryon threshold channel $mB$,
with a hamiltonian matrix: 
\begin{equation}
\lbrack H]\simeq \left( 
\begin{array}{cc}
M_{m}+M_{B} & a \\ 
a^{\ast } & M_{3q^{2b}}%
\end{array}%
\right) 
\end{equation}%
where $M_{3q^{2b}}$ stands for the mass of the $3q^{2b}$ state, $M_{m}$ and $%
M_{B}$ for the masses of the meson and baryon respectively and $a$ for a
fitting parameter giving account of the interaction.

In order to proceed to calculate the eigenvalues we have to choose a
particular $3q^{2b}$ model and establish a criterion for the choice of the $%
mB$ channel for each anomaly. We shall use as $M_{3q^{2b}}$ the values
calculated in Ref.~\cite{Cap86}. As $mB$ we shall take for granted the $N%
\overline{K}$ channel for $\Lambda _{S_{01}}(1405)$ and the $\sigma N$
channel for $N_{P_{11}}(1440)$. For $\Delta
_{D_{35}}(1930),$ $\Delta _{D_{33}}(1940)$ and $\Delta _{S_{31}}(1900)$ we
shall select $\rho \Delta $ (equivalently we could have preferred the almost
degenerate $\omega \Delta )$ as suggested by our phenomenological analysis.
For the same reason $\pi N_{D_{15}}(1675)$ will be employed for $\Delta
_{F_{35}}(\sim 1720).$ For $\Delta _{P_{31}}(1750)$ we shall use $\pi
N_{S_{11}}(1650)$ and for $\Delta_{P_{33}}(1600)$ we shall take
$\pi N_{P_{13}}(1520)$.
\begin{table}[b]
\begin{tabular}{|c|cc|cc|c|c|}
\hline
PDG Resonance & 
$mB$ threshold & 
Prob. & 
$3q^{2b}$ & 
Prob. & 
$M$ & 
Experiment \\ \hline\hline
$\Delta_{P_{33}}(1600)(***)$ & [$\pi\,N_{P_{13}}(1520)$] (1660) & 81.1\% & 
1795 & 18.9\% & 1619 & 1550--1700 \\ \hline
$N_{P_{11}}(1440)(****)$ & [$\sigma\,N$] (1540) & 50.0\% & 1540 & 50.0\% & 
1455 & 1420--1470 \\ \hline
$\Delta_{D_{35}}(1930)(***)$ &  & 83.4\% & 2155 & 16.6\% & 1964 & 1900--2020
\\ 
$\Delta_{D_{33}}(1940)(*)$ & [$\rho\,\Delta]$ (2002) & 82.2\% & 2145 & 17.8\%
& 1962 & 1840--2040$^\dagger$ \\ 
$\Delta_{S_{31}}(1900)(*)$ &  & 81.5\% & 2140 & 18.5\% & 1961 & 1850--1950
\\ \hline
$\Delta_{P_{31}}(1750)(*)$ & [$\pi\,N_{S_{11}}(1650)$] (1790) & 62.8\% & 1835
& 37.2\% & 1725 & 1710--1780 \\ \hline
$\Delta_{F_{35}}(\approx 1720)$(N.C.) & [$\pi\,N_{D_{15}}(1675)$] (1815) & 
74.4\% & 1910 & 25.6\% & 1765 & 1660--1785$^{\dagger \dagger}$ \\ \hline
$\Lambda_{S_{01}}(1405)(****)$ & [$\bar K\,N$] (1434) & 78.2\% & 1550 & 
21.8\% & 1389 & 1400--1410 \\ \hline
\end{tabular}%
\caption{Predicted masses, $M$, for the anomalies as compared to
experimental data from Ref.~\protect\cite{PDG06}, Ref.~\protect\cite{Cut80}
(indicated by the superindex $\dagger$), and Ref.~\protect\cite{Vra00}
(indicated by the superindex $\dagger\dagger$). Two-body quark-model masses (%
$3q^{2b}$) are taken from Ref.~\protect\cite{Cap86}. Probabilities (Prob.)
for meson-baryon and $3q$ components are also shown. All masses are in MeV.}
\label{t2}
\end{table}

\begin{figure}
\vspace*{9cm}
  \includegraphics[width=11cm]{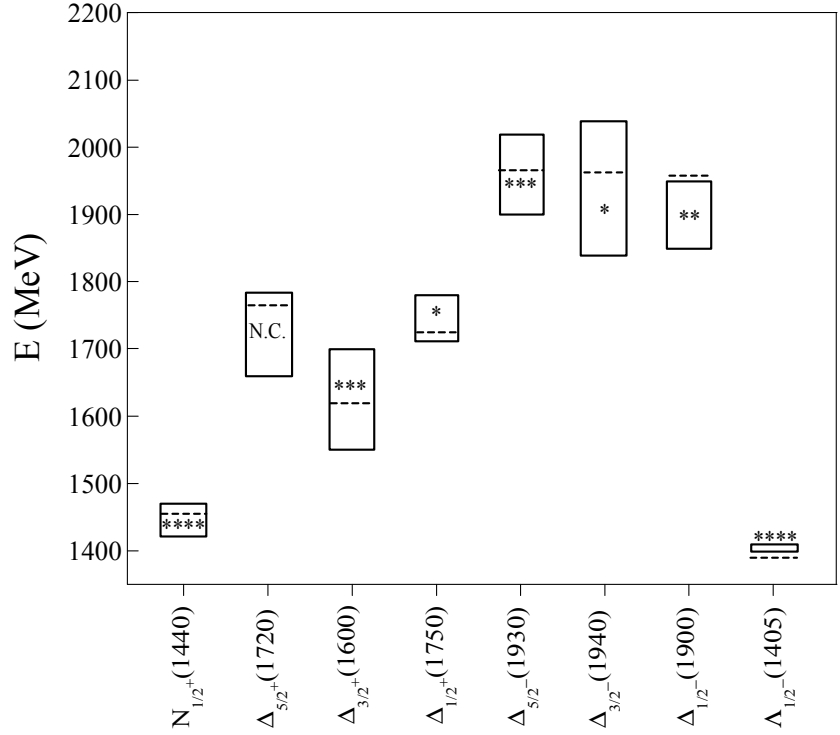}
  \vspace{-10cm}
\caption{Predicted masses for the anomalies (dashed lines) as compared to
the experimental mass intervals detailed in Table~\protect\ref{t2} (boxes).
N.C. means non-cataloged resonance.}
\label{fig3}
\end{figure}

Although the value of $\left\vert a\right\vert $ might vary depending on the
configurations involved in each $(mB)-3q$ coupling we shall use for the sake
of simplicity the same value in all cases. The $M$ results for $%
\left\vert a\right\vert =85$ MeV are numerically detailed in Table~\ref{t2}
where the values for $M_{3q^{2b}}$ and for ($M_{m}+M_{B})$ in the chosen $mB$
channel as well as their probabilities to give $M$ are also displayed.
As can be checked the improvement of the description with respect to $%
3q^{2b} $ is astonishing. All the predicted $M$ masses lye very close to
the PDG average masses for the anomalies. In Fig.~\ref{fig3} the $M$
values for $\left\vert a\right\vert =85$ MeV are drawn\ as compared to the
experimental mass intervals.

We interpret these results as providing strong quantitative support to our
former qualitative description of the anomalies. Regarding their nature a
look at the probabilities reveal they are mostly meson-baryon states.
Actually a meson-baryon probability greater or equal than $50\%$ can serve
as a criterion to identify an anomaly. Nonetheless the coupling to the $3q$
component is essential to lower their masses making them more stable against
decay into $m+B$. 

It should be emphasized that similar results could be obtained for any other
spectroscopic $3q^{2b}$ model through a fine tuning of the value of $%
\left\vert a\right\vert $ (note that the small value of $\left\vert
a\right\vert $ as compared to $M_{3q^{2b}}$ and $(M_{m}+M_{B})$ provides an 
\textit{a posteriori} validation of our method)$.$ This comes from the
expression of the eigenvalues where it is clear that even for $\left\vert
a\right\vert =0$ one gets $M=M_{m}+M_{B}$ which according to our $mB$
choice is much closer to the PDG mass of the anomaly than $M_{3q^{2b}}$. 
This means that concerning the mass of the anomalies the
coupling of meson-baryon to $3q$ components may play the role of a general 
\emph{healing mechanism} for spectroscopic models.

\section{Summary}
\label{sec5}

We have studied the heavy baryon spectra by means of the
Faddeev method in momentum space. These results should be highly
valuable both from the theoretical and experimental points of 
view. Theoretically, it should be a powerful tool
for testing different approximate methods to solve the 
three-body problem. Experimentally, the remarkable 
agreement with known experimental data make our
predictions highly valuable as a guideline to experimentalists.

Heavy baryons constitute an extremely interesting problem
joining the dynamics of light-light and heavy-light subsystems
in an amazing manner. While the mass difference between members
of the same SU(3) configuration, either 
$\mathbf{\bar{3}_F}$ or $\mathbf{6_F}$, is determined
by the perturbative one-gluon exchange, the mass difference
between members of different representations comes mainly
determined by the dynamics of the light diquark, and should
therefore be determined in consistency with the light 
baryon spectra. There is therefore a remnant effect
of pseudoscalar forces in the two-light quark subsystem. 

For light baryons we propose that $4q1\overline{q}$ components, in the form of $S$
wave meson-baryon channels which we identify, play an essential role in the
description of the anomalies, say baryon resonances very significantly
overpredicted by three-quark models based on two-body interactions. As a
matter of fact by considering a simplified description of the anomalies as
systems composed of a free meson-baryon channel interacting with a
three-quark confined component we have shown they could correspond mostly to
meson-baryon states but with a non-negligible $3q$ state probability which
makes their masses to be below the meson-baryon threshold. The remarkable
agreement of our results with data in all cases 
suggests the implementation of meson-baryon threshold 
effects as an essential physical mechanism to give account of spectral 
states poorly described by constituent quark models.

\begin{theacknowledgments}
This work has been partially funded by the Spanish MEC and EU FEDER 
under Contract No. FPA2007-65748, by EIII 506078,
by JCyL under Contract No. SA016A17, and by COFAA-IPN (M\'exico).
\end{theacknowledgments}

\end{document}